\documentclass[aps,prl,twocolumn,floatfix,superscriptaddress,showpacs,amsfonts,amssymb,amsmath,preprintnumbers]{revtex4}
\usepackage{bm}
\usepackage{epsfig}
\usepackage[dvips]{color}
\usepackage{graphicx}
\usepackage{color}
\usepackage{subfigure}

\newcommand{\fig}[1]{Fig.~\ref{#1}}

\begin{document}

\title{Formation of Plasmoid Chains in Magnetic Reconnection}

\author{R.\ Samtaney}
\affiliation{Princeton Plasma Physics Laboratory, Princeton University, Princeton, New Jersey 08543}
\author{N.\ F.\ Loureiro}
\affiliation{EURATOM/UKAEA Fusion Association, Culham Science Centre, 
Abingdon, OX14 3DB, UK}
\author{D.\ A.\ Uzdensky}
\affiliation{Department of Astrophysical Sciences/CMSO, 
Princeton University, Princeton, New Jersey 08544}
\author{A.\ A.\ Schekochihin}
\affiliation{Rudolf Peierls Centre for Theoretical Physics, University of Oxford, Oxford OX1 3NP, UK}
\author{S.\ C.\ Cowley}
\affiliation{EURATOM/UKAEA Fusion Association, Culham Science Centre,
 Abingdon, OX14 3DB, UK}
\affiliation{Plasma Physics, Blackett Laboratory, Imperial College, London~SW7~2AZ, 
UK}
\date{\today}

\begin{abstract}
A detailed numerical study of magnetic reconnection in resistive 
MHD for very large, previously inaccessible, 
Lundquist numbers ($10^4\le S\le 10^8$) is reported.
Large-aspect-ratio Sweet-Parker current sheets are 
shown to be unstable to super-Alfv\'enically 
fast formation of plasmoid (magnetic-island) chains. 
The plasmoid number scales as $S^{3/8}$ and 
the instability growth rate in the linear stage as $S^{1/4}$, 
in agreement with the theory by Loureiro et al.\ [Phys.\ Plasmas~{\bf 14}, 100703 (2007)]. 
In the nonlinear regime, plasmoids continue to grow faster 
than they are ejected and completely disrupt the reconnection layer. 
These results suggest that high-Lundquist-number reconnection is 
inherently time-dependent and hence call for a substantial 
revision of the standard Sweet-Parker quasi-stationary picture for $S>10^4$.
\end{abstract}

\pacs{52.35.Vd, 94.30.Cp, 96.60.Iv, 52.35.Py, 52.65.Kj}

\maketitle


\paragraph{Introduction.} 
Magnetic reconnection is a fundamental plasma process of rapid rearrangement
of magnetic field topology, accompanied by a violent
release of magnetically-stored energy and its conversion
into heat and into non-thermal particle energy. 
It is of crucial importance for numerous physical phenomena 
such as solar flares and coronal mass ejections \cite{sweet_review,yokoyama_01}, 
magnetic storms in the Earth's magnetosphere \cite{dungey_61,bhatta_04,xiao_06}, 
and sawtooth crashes in tokamaks \cite{yamada_94,hastie_97}.
Reconnection times in these environments are observed to be very short,
usually only 10 to 100 times longer that the global Alfv\'en transit time, 
$\tau_A = L/v_A$, where $L$ is the characteristic system size
and $v_A$ is the Alfv\'en speed.
This is in direct contradiction with the classical 
Sweet-Parker (SP) \cite{sweet,parker} reconnection
model, which employs the simplest possible non-ideal description of 
plasma --- two-dimensional (2D) 
resistive magnetohydrodynamics (MHD) --- and predicts a very long 
reconnection time scale $\tau_{\rm rec}\sim\tau_A S^{1/2}$, where $S=Lv_A/\eta\gg 1$ is the
Lundquist number, and $\eta$ is the resistivity (or magnetic diffusivity) 
of the plasma.
Both numerical simulations \cite{biskamp_86, uzdensky_00} 
and laboratory experiments \cite{ji_99} have confirmed the SP theory
for collisional plasmas where resistive MHD with smoothly varying
(e.g., Spitzer) resistivity must apply. 
Because of this discrepancy between the MHD picture and observations,
efforts to understand magnetic reconnection have moved beyond simple 
resistive MHD to increasingly sophisticated and realistic plasma-physics 
frameworks incorporating 
collisionless processes such as anomalous resistivity or two-fluid effects, 
where, indeed, fast reconnection rates have been found \cite{ugai_77,birn_01}.  
For these reasons, simple resistive-MHD reconnection has come to be viewed as
well understood, uninteresting, and mostly irrelevant.

However, most previous numerical studies of 
resistive MHD reconnection have been limited by resolution constraints
to relatively modest Lundquist numbers ($S\sim10^4$). 
The same is true for dedicated reconnection experiments. 
On the other hand, in most real applications of reconnection, 
the Lundquist numbers are much larger (e.g., $S\sim10^{12}$ 
in the solar corona, $S\sim 10^8$ in large tokamaks). 
Thus, there is a large gap between 
the extreme parameter regime that we would like to understand and that 
accessible to the numerical and experimental studies performed so far. 
Asymptotically high values of $S$ have never been probed by numerical simulations 
and, therefore, one cannot really claim a complete understanding of magnetic 
reconnection even in the simplest framework of 2D MHD with a (quasi-)uniform resistivity.

Of particular interest is the possibility 
that current sheets with large aspect ratios 
$L/\delta_{\rm SP}\sim S^{1/2}$, where $\delta_{\rm SP}$ is the width of the SP current layer, 
should be unstable and break up into chains of secondary magnetic islands, 
or plasmoids --- a phenomenon absent from the SP theory. 
A tearing instability of large-aspect-ratio current sheets was 
anticipated by Bulanov et al.\ \cite{bulanov_79} and Biskamp \cite{biskamp_86}. 
Current sheets were, indeed, found to be unstable in those numerical experiments
where $S\sim10^4$ was reached 
(e.g.\ \cite{biskamp_86,lee_fu,jin_91,loureiro_05,lapenta_08,bhatta_09}). 
Current-sheet instability and plasmoid formation have 
also been observed in numerical reconnection studies using other physical
descriptions, e.g., fully kinetic 
simulations \cite{daughton_06,daughton_09,drake_plasmoids_06}, 
and there is tentative observational evidence \cite{lin_08,bemporad_08} 
that they might play a key role in the dynamics of magnetic
reconnection in the Earth magnetosphere and in solar flares.
In fusion devices, plasmoid formation is less well diagnosed but recent 
results from the TEXTOR tokamak \cite{liang_07} suggest 
that they might also be present.
Theoretically, plasmoid formation has been proposed as a mechanism of 
fast reconnection \cite{shibata_01,lapenta_08} and 
non-thermal particle acceleration in reconnection events \cite{drake_plasmoids_06}.
Thus, plasmoids seem to be as ubiquitous as magnetic reconnection itself. 
However, even though their appearance has been reported by many authors, 
neither the plasmoid formation in the limit of asymptotically large $S$  
nor its effect on the reconnection process have been systematically
investigated on any quantitative level and remains poorly understood.

As the first step towards this goal, Loureiro et al.\ \cite{loureiro_07}  
developed a linear theory of the instability of large-aspect-ratio current 
sheets that, unlike in the calculation of Ref.~\cite{bulanov_79}, emerges 
from a controlled asymptotic expansion in large $S$. 
Mathematically, the instability resembles a tearing instability 
with large $\Delta'$, leading to the formation of 
an inner layer with the width $\delta_{\rm inner}\sim S^{-1/8}\delta_{\rm SP}$. 
The instability is super-Alfv\'enically fast, 
with the maximum growth rate scaling as $\gamma\tau_A\sim S^{1/4}$; 
the fastest-growing mode occurs on a 
scale that is small compared to the length of the current sheet, viz., 
the number of plasmoids formed along the sheet scales as $S^{3/8}$. 

In this Letter, we report the next logical step towards the detailed 
assessment of the role of plasmoids in magnetic reconnection:
the first numerical evidence that not only do current sheets go 
unstable but that they do so in the extremely fast way predicted 
by Ref.~\cite{loureiro_07} and the instability quantitatively 
obeys the scalings derived there. To this end, we perform a set of 
2D MHD simulations of an SP reconnection layer 
with uniform resistivity and asymptotically large Lundquist numbers 
$10^4\le S\le10^8$. 

\paragraph{Numerical Set Up.} 
Probing such previously unattainable values of the Lundquist number 
is made possible by a special numerical set~up that effectively zooms 
in on the SP current sheet by choosing a simulations box 
whose size in the direction across the reconnection layer ($x$) 
is just somewhat larger than, but generally tied to,  
the SP thickness, $L_x \gtrsim \delta_{\rm SP}$, 
while in the direction along the layer ($y$), the box 
covers a finite fraction of the global length $L$ of the current sheet.
The boundary conditions are used to 
mimic the asymptotic matching between the global and local solutions 
(in the spirit of~\cite{uzdensky_00}). Let us explain how this is done. 

We solve the standard set of compressible resistive 
MHD equations (the adiabatic index is $5/3$; viscosity and thermal 
conductivity are ignored) in an elongated 2D box, 
$[-L_x,L_x]\times[-L_y,L_y]$. 
At the upstream boundaries ($x=\pm L_x$), we prescribe the 
reconnecting component of the magnetic field,
$B_y(x=\pm L_x,y)=\pm B_{\rm in}$
and the incoming velocity, $v_x(x=\pm L_x,y)=\mp v_{\rm in}$. 
As the box is understood to model an SP current sheet, 
we set $L_x=\delta_{\rm SP}=LS^{-1/2}=(L\eta/v_A)^{1/2}$,
where $v_A$ is the Alfv\'en speed corresponding to $B_{\rm in}$
and $L$ is the (half-)length of the current sheet. 
We should then have $v_{\rm in}=v_A S^{-1/2}=(v_A\eta/L)^{1/2}$. 
We choose our code units so that $v_A=1$ and $L=1$. 
Then setting $L_x=\eta^{1/2}$ and $v_{\rm in}=\eta^{1/2}$ enforces 
a fixed SP reconnection rate based on $v_A=1$ and $L=1$. 
Choosing $L_y=1$ would 
correspond to simulating the entire length of the current sheet, but it is clear 
that in this local set~up only an inner fraction of the sheet can be computed 
accurately, so we choose $L_y=0.24$. At the downstream boundaries $y=\pm L_y$, 
free outflow boundary conditions are imposed. 
The method of characteristics is used to determine the remaining boundary conditions.
Compressibility effects are minimized by ensuring that the Mach number 
$M=v_A/c_s$, where $c_s$ is the sound speed, is small 
(in our simulations, it never exceeds $0.1$).
The initial conditions are chosen so as to represent qualitatively an SP-like 
current layer (using the Harris sheet profiles). 
We do not choose an initial perturbation with a particular wave number; 
instead the system itself is allowed to pick the most unstable wave number. 

\paragraph{Time Evolution of the Instability.} 
\begin{figure*}[t]
\parbox{\textwidth}{
\subfigure
          {
          \includegraphics[angle=90,width=1.0\textwidth,height=.5cm]{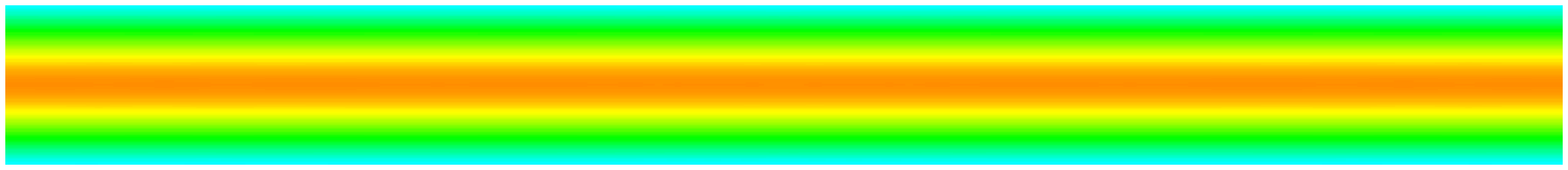}}
\subfigure
          {
          \includegraphics[angle=90,width=1.0\textwidth,height=.5cm]{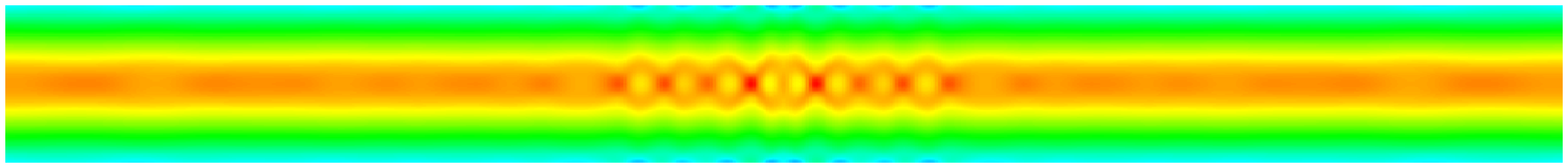}}
\subfigure
          {
          \includegraphics[angle=90,width=1.0\textwidth,height=.5cm]{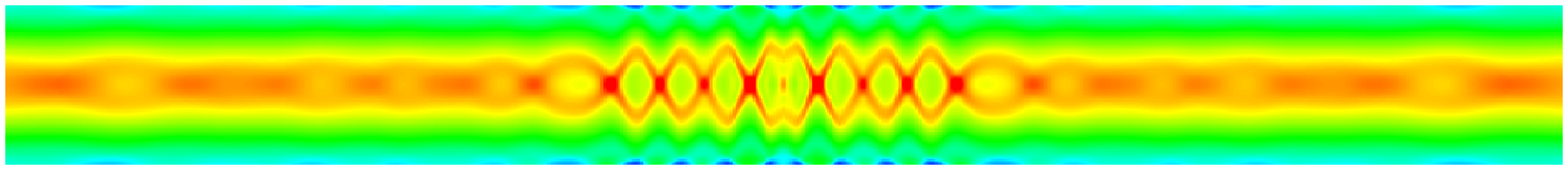}}
\subfigure
					{
          \includegraphics[angle=90,width=1.0\textwidth,height=.5cm]{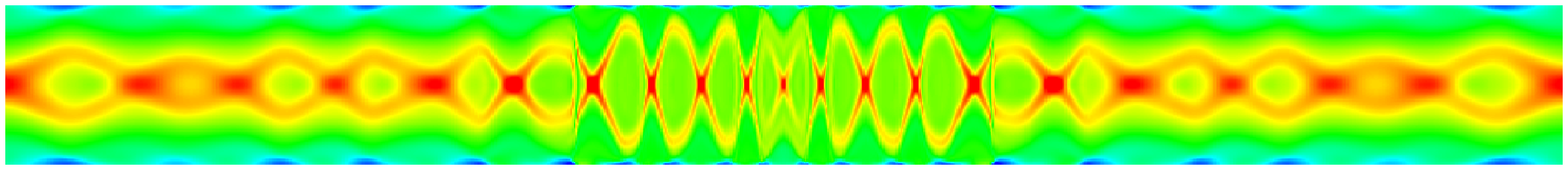}}
}
\caption{\label{fig:JzS7} Contour plots of the current density showing the time 
evolution of an SP current sheet for $S=10^7$. The times shown 
are, from top to bottom,  $t=0.63\tau_A$, $t=0.96\tau_A$, $t=1.09\tau_A$ and $t=1.27\tau_A$.
The domain shown is $-\delta_{\rm SP}\le x\le\delta_{\rm SP}$ 
(inflow direction, vertical), and $-0.12 L\le y\le 0.12 L$ 
(outflow direction, horizontal), where $\delta_{\rm SP}\simeq3\times10^4$ 
is the SP layer width and $L=1$ is the (half-)length of the current sheet
(see text; only the central half of the simulation box is shown).}
\end{figure*}
\begin{figure*}
\parbox{\textwidth}{
\subfigure
           {
           \includegraphics[angle=90,width=1.0\textwidth,height=.5cm]{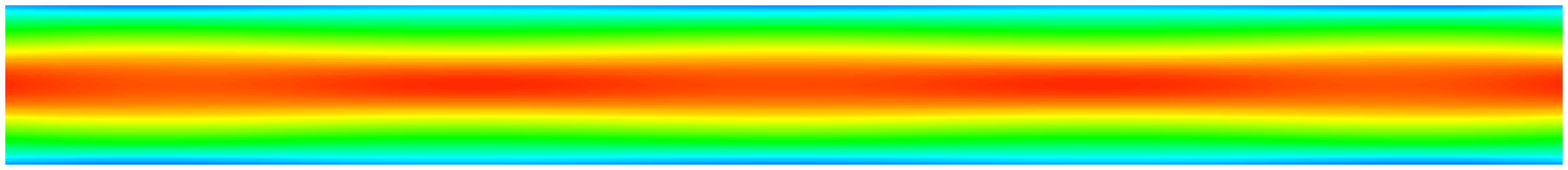}}
\subfigure
           {
           \includegraphics[angle=90,width=1.0\textwidth,height=.5cm]{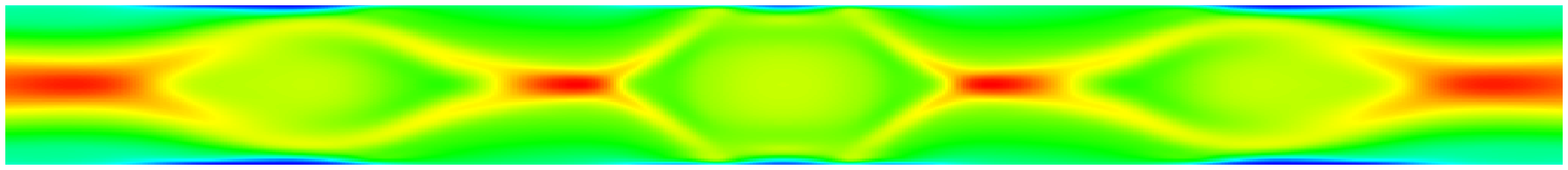}}
\subfigure
           {
           \includegraphics[angle=90,width=1.0\textwidth,height=.5cm]{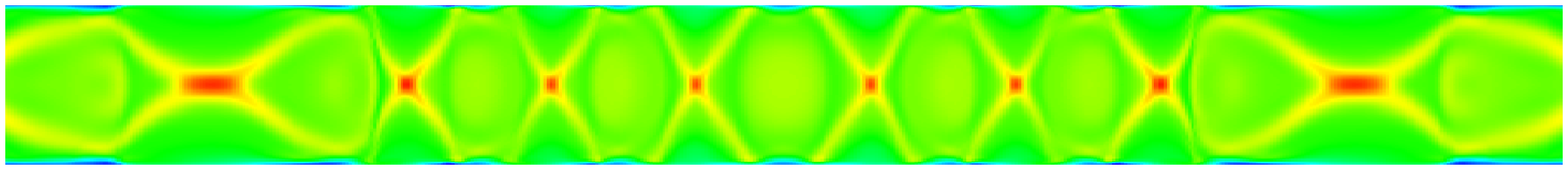}}
}
\caption{\label{fig:Psi} Contour plots of the current density for 
(from top to bottom) $S=10^4$ ($t=4.5\tau_A$), $S=10^5$ ($t=2.9\tau_A$) 
and $S=10^6$ ($t=2.6\tau_A$); the case $S=10^7$ is shown in \fig{fig:JzS7}.}
\end{figure*}

\begin{figure}[b]
\parbox{\columnwidth}{
\includegraphics[width=0.6\columnwidth,angle=270]{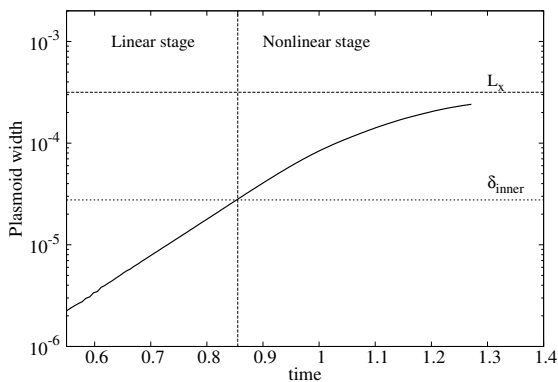}
}           
\caption{Time evolution of the half-width of the plasmoid closest to the center 
of the sheet for $S=10^7$. 
Half-widths of the inner layer, $\delta_{\rm inner}$ \cite{loureiro_07}, and 
of the SP sheet, $\delta_{\rm SP}=L_x$, are shown for reference.}
\label{fig:plas_width}
\end{figure}

\begin{figure}[b]
\parbox{\columnwidth}{
\includegraphics[width=0.6\columnwidth,angle=270]{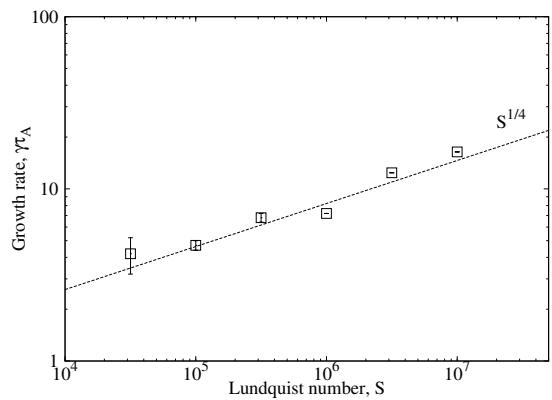}
}           
\caption{The growth rate of the plasmoid closest to the center of the sheet vs.\ $S$. 
The theoretical slope $S^{1/4}$ \cite{loureiro_07} is shown for reference. 
For $S>10^{7}$, we could not calculate the growth rates accurately 
because the linear regime was too short lived.}
\label{fig:plas_grate}
\end{figure}

For $S<10^4$, the current density at $x=0$ settles down to
a quasi-steady state, and no plasmoids are observed, 
consistent with SP theory. 
As the Lundquist number is increased, this picture changes dramatically.
The system does not settle into a steady state --- 
instead, as predicted by the linear theory \cite{loureiro_07}, 
the layer becomes unstable and secondary islands (plasmoids) form, 
with reconnection occurring at multiple $X$-points. 
This is illustrated in \fig{fig:JzS7}, where the time evolution of
the current density for $S=10^7$ is shown. 
We see that a plasmoid chain develops along the sheet and 
also that the plasmoids closer to the center of the sheet grow 
faster than those farther away from it. 
In \fig{fig:plas_width}, we show the time evolution of the 
width of the plasmoid closest to the center of the sheet. 
At early times, it grows exponentially. 
The growth rates for different values of $S$ are
plotted in \fig{fig:plas_grate}. The scaling $\gamma\tau_A\sim S^{1/4}$ 
predicted by the linear theory \cite{loureiro_07} is obeyed extremely 
well. The fact that the growth rate for the off-center plasmoids 
is a decreasing function of the distance along the sheet was not 
captured in the simple equilibrium model used in Ref.~\cite{loureiro_07} 
but it does emerge in the calculation for a general SP 
equilibrium \cite{loureiro_09}.

\begin{figure}[t]
\parbox{\columnwidth}{
\includegraphics[width=0.6\columnwidth,angle=270]{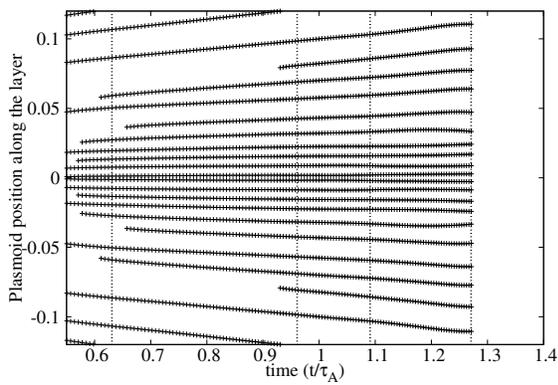}
}
\caption{Plot of the position of the $O$ points vs.\ time for $S=10^{7}$.
Dotted lines mark the times at which the current sheet is shown in \fig{fig:JzS7}.}
\label{fig:plas_pos}
\end{figure}

The linear stage ends when the plasmoid width exceeds the width of the inner 
layer, $\delta_{\rm inner}\sim S^{-1/8}\delta_{\rm SP}$ (see \fig{fig:plas_width}). 
The subsequent nonlinear growth is slower, but still sufficiently rapid to permit 
the plasmoids to reach the width of the current sheet before being advected out 
by the mean outflow along the SP sheet, $v_y\sim (y/L)v_A$. 
That this outflow causes the plasmoids to drift outwards 
is illustrated by \fig{fig:plas_pos}, where we plot the 
time traces of the plasmoid $O$ points along the sheet.  
The plasmoids that are further away from the center of the sheet move faster 
due to faster outflow. 
Note that new $O$-point traces appear in between already existing ones --- 
we interpret this as evidence of secondary-plasmoid generation, arising from 
the collapse of the $X$ points between the primary plasmoids into secondary current 
sheets, which then go unstable (cf.\ \cite{loureiro_05}). 
The breaking of the secondary sheets 
is also visible in the last two frames of \fig{fig:JzS7}. 
In some cases, we have also observed coalescence of nearby plasmoids (not shown). 

Once the plasmoids grow enough to touch the wall,  
the simulation becomes invalid and is stopped. The fact that they do touch the wall 
means that in a more general global set~up, they would grow to exceed 
the SP width $\delta_{\rm SP}$, so the current sheet is broken up. 

\begin{figure}[b]
\parbox{\columnwidth}{
\includegraphics[width=0.6\columnwidth,angle=270]{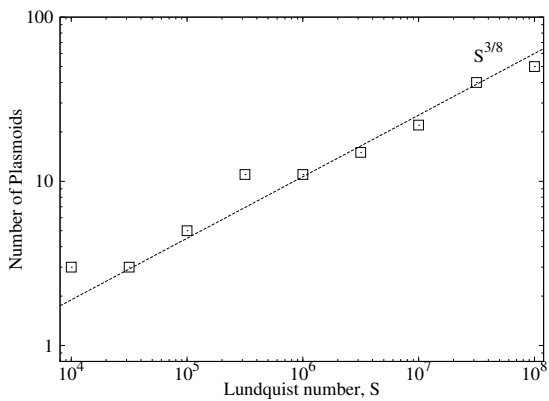}
}           
\caption{The maximum number of plasmoids in the central 
part of the sheet, $-0.12L\le y\le 0.12L$, vs.\ $S$.
The theoretical slope from linear theory, $S^{3/8}$ \cite{loureiro_07}, 
is shown for reference.}
\label{fig:plas_number}
\end{figure}

\paragraph{Spatial Structure of the Plasmoid Chain.}
The development of multiple magnetic islands for different values of $S$ is 
illustrated in \fig{fig:Psi}: an increasing number of plasmoids is observed
as the Lundquist number increases. \fig{fig:plas_number} shows the 
number of plasmoids vs.\ $S$. Again, there is excellent agreement with the 
scaling $\sim S^{3/8}$ of the most unstable wave number predicted by the 
linear theory \cite{loureiro_07}. 
Note that to avoid possible boundary-condition effects, we only count the 
number of plasmoids present in the central half of the simulation domain,  
$-0.12L<y<0.12L$. 

Note that the decrease in the linear growth rate with $y$ means that in practice 
the simulation must be run into a nonlinear state to observe the formation of the 
entire plasmoid chain. However, by the time the outer plasmoids become detectable 
(but still linear), the more central plasmoids can already be well into their 
nonlinear evolution, so some secondary plasmoid generation 
(and, in some cases, coalescence) might have already taken place. 
This means that there is a degree of imprecision in measuring the number 
of the primary plasmoids resulting from the linear instability of the 
entire current sheet (the diagnostic in \fig{fig:plas_number} counts 
the maximum number of plasmoids during the run), 
especially at the largest values of $S$. 
However, the errors this introduces 
in \fig{fig:plas_number} are not large and do not affect the validity 
of our claim that the theoretical scaling $S^{3/8}$ is obeyed. 
At even larger Lundquist numbers $S>10^8$, the dynamics along the sheet 
is likely to be even messier, so a statistical description of plasmoid 
chains may be required (cf.\ \cite{shibata_01,lapenta_08}).
 
\paragraph{Discussion.}
The key questions that remain to be answered are how large the plasmoids 
grow after they exceed the width of the SP layer, 
and what is their impact on the overall reconnection rate.
Addressing these questions requires global (or, at least, intermediate-scale)
simulations capturing regions both interior and exterior to the current sheet. 
We could not computationally afford 
such simulations and simultaneously resolve the very 
large values of $S$ required to diagnose the current-sheet instability 
in its asymptotic form. In our view, it was important, before undertaking 
a global reconnection study, to understand the nature of the instability. 
The conclusion of this Letter is that the instability 
exists, is super-Alfv\'enically fast (cf.\ \cite{daughton_09,bhatta_09}) 
and, in the limit of large $S$, 
quantitatively follows the linear theory of Ref.~\cite{loureiro_07,loureiro_09}. 
We stress that this is the first numerical study that has been able 
to make this statement and thus, in a sense, demystify the phenomenon 
of multiple-plasmoid generation. This conclusion puts further investigations
of the plasmoid effect on magnetic reconnection on a firm theoretical footing. 

The body of numerical evidence for plasmoids is now so overwhelming that 
there should remain little doubt of their importance in reconnection processes.
It is clear that for sufficiently large systems,  
plasmoid-dominated reconnection layers are inevitable. 
Plasmoid formation and magnetic reconnection are thus 
inextricably linked and further progress in understanding reconnection 
in realistic systems necessarily requires a theory that takes 
the plasmoid dynamics into account. It also requires experimental 
evidence --- however, present-day magnetic reconnection experiments have not yet been 
able to observe plasmoid formation, most likely because of the moderate current sheet aspect 
ratios. The effect of plasmoids on reconnection is, in our view, one of the 
natural objects of emphasis for the next generation
of dedicated magnetic reconnection experiments.

\begin{acknowledgments}
R.S. was supported by USDOE Contract DE-AC020-76-CH03073. 
D.A.U.\ was supported by NSF Grant PHY-0215581 (PFC: CMSO).
A.A.S.\ was supported in part by an STFC Advanced Fellowship and 
STFC Grant ST/F002505/2. This work was supported in part by EPSRC 
and the European Commission under the contract of Association between 
EURATOM and UKAEA. The views and opinions expressed herein do not necessarily 
reflect those of the European Comission. R.S.\ and D.A.U.\ thank the Leverhulme 
Trust International Network for Magnetised Plasma Turbulence for travel support.
\end{acknowledgments}

\bibliography{plasmoids_PRL_submit}

\end{document}